\documentclass[10pt]{article}
\pdfoutput=1
\usepackage{chicago,epic,graphpap,graphics,float,tabularx,paralist,longtable,fancyhdr,fancyvrb,bbm}
\usepackage{amsmath}
\usepackage{amstext}
\usepackage{amsbsy}
\usepackage{amsthm}
\usepackage{amsfonts}
\usepackage{amssymb}
\usepackage{amscd}
\usepackage{eucal}
\def\ito{It{\^o}}
\def\X{{\overline X}}
\def\V{{\bf V}}
\def\VV{\widetilde{\bf V}}
\def\1{{\mathbbm 1}}
\def\E{{\mathbb E}}
\def\eqdef{\triangleq}
\def\sumi{\sum_{i=1}^n}
\def\s{\sigma}

\begin{document}

\centerline{\Large\bf Identification of Atlas models}
\vspace{5pt} \centerline{\large Robert Fernholz\footnote{INTECH, One Palmer Square, Princeton, NJ 08542.  bob@bobfernholz.com.}} \centerline{ \today}
\vspace{10pt}
\begin{abstract}
Atlas models are systems of \ito\ processes with parameters that depend on rank. We show that the parameters of a simple Atlas model can be identified by measuring the variance of the top-ranked  process for different sampling intervals.
\end{abstract}
\vspace{10pt}

Let $X_1,\ldots,X_n$ be an Atlas model with 
\begin{equation}\label{1.1}
dX_i(t) = \sumi g_k\1_{\{X_i(t)=X_{(k)}(t)\}}dt + \s\,dW_i(t),
\end{equation}
where $\s^2>0$, the $g_k$ are constants such that $g_1+\cdots+g_n=0$ and $g_1+\cdots+g_m<0$ for $m<n$, and $(W_1,\ldots,W_n)$ is an $n$-dimensional Brownian motion (see. e.g., \citeN{F:2002} or \shortciteN{BFK:2005}). Let $X_{(1)}(t)\ge\cdots\ge X_{(n)}(t)$ represent the ranked processes $X_i(t)$. The {\em depth} of an Atlas model is the number of processes in it, in this case, $n$. Let us assume that the gap processes $X_{(k)}-X_{(k+1)}$ are in their steady-state distributions with initial values such that $X_1(0)+\cdots+X_n(0)=0$. (Here we need the existence of steady-state distributions for the gap processes, but we don't seem need the fact that these distributions are exponential --- in certain cases, at least --- as is proved in \citeN{IK:2010} and \shortciteN{BFK:2005}.)

Let $Z(t)$ be a continuous semimartingale defined for $t\ge0$ such that $\E\big[\big(Z(s+t)-Z(s)\big)^2\big]$ exists for $s,t\ge0$ and is independent of $s$. Then the {\em variogram} $\V_Z$ of $Z$ is the real-valued function defined for $t> 0$ by
\begin{equation}\label{1}
\V_Z(t) \eqdef \E\bigg[\frac{\big(Z(t)-Z(0)\big)^2}{t}\bigg].
\end{equation}
(This may be an unusual definition of variogram, but it is convenient in our context and contains the same information as the classical version.) For a Brownian motion $B$, the variogram is $\V_B(t)\equiv 1$. Here we consider the variogram $\V_{X_{(1)}}$ of the top-ranked process in an Atlas model. 

Since the $g_k$ add up to zero, the average $\X$  of the $X_i$ will be
\begin{equation*}
	\X(t) = \frac{1}{n}\sumi X_i(t)
	=  \frac{\s}{n}\sumi W_i(t)
	= \frac{\s}{\sqrt{n}}\,W(t),
\end{equation*}
where $W$ is a new Brownian motion. Hence, for $t>0$,
\begin{equation}\label{2}
\E\bigg[\frac{\X^2(t) }{t}\bigg]= \frac{\s^2}{n}\E\bigg[\frac{ W^2(t)}{t}\bigg]= \frac{\s^2}{n}.
\end{equation}

Since the gap processes $X_{(k)}-X_{(k+1)}$ are assumed to be in their steady-state distribution, we have
\begin{equation*}
\E\big[\big(X_{(1)}(t)-\X(t)\big)^2\big] = \E\big[\big(X_{(1)}(0)-\X(0)\big)^2\big] = C,
\end{equation*}
for a constant $C>0$. Now,
\begin{align}
\E\big[X^2_{(1)}(t)\big]&=\E\big[\big(\X(t)+\big(X_{(1)}(t)-\X(t)\big)\big)^2\big]\notag\\
&=\E\big[\X^2(t)\big] + 2\E\big[\X(t)\big(X_{(1)}(t)-\X(t)\big)\big]+\E\big[\big(X_{(1)}(t)-\X(t)\big)^2\big]\notag\\
&= \E\big[\X^2(t)\big] + 2\E\big[\X(t)\big(X_{(1)}(t)-\X(t)\big)\big]+C,\label{4}
\end{align}
and, by the Cauchy-Schwarz inequality,
\begin{align*}
\Big(\E\big[\X(t)\big(X_{(1)}(t)-\X(t)\big)\big]\Big)^2&\le \E\big[\X^2(t)\big]\E\big[\big(X_{(1)}(t)-\X(t)\big)^2\big],\\
&=C\E\big[\X^2(t)\big].
\end{align*}
Hence,
\begin{equation*}
\lim_{t\to\infty}\frac{\E\big[\X(t)\big(X_{(1)}(t)-\X(t)\big)\big]}{t}=0,
\end{equation*}
so \eqref{2} and \eqref{4} imply that
\begin{equation}\label{7.6}
\lim_{t\to\infty}\E\bigg[\frac{X_{(1)}^2(t)}{t}\bigg]=\frac{\s^2}{n}.
\end{equation}

In terms of variograms, \eqref{7.6} becomes
\begin{equation*}
\lim_{t\to\infty}\V_{X_{(1)}}(t) =\frac{\s^2}{n},
\end{equation*}
and from  \eqref{1.1} we have
\begin{equation*}
\V_{X_{(1)}}(0)\eqdef\lim_{t\downarrow0}\V_{X_{(1)}}(t) =\s^2.
\end{equation*}
Hence, the {\em relative variogram} of $X_{(1)}$,
\begin{equation*}
\VV_{X_{(1)}}(t)\eqdef \frac{ \V_{X_{(1)}}(t)}{\V_{X_{(1)}}(0)},
\end{equation*}
satisfies
\begin{equation}\label{9}
\lim_{t\to\infty} \VV_{X_{(1)}}(t) = \frac{1}{n},
\end{equation}
and this provides an estimator for the depth, $n$, of the model.

There are a couple of simple manipulations of the model \eqref{1.1} that are worth noting. First, if we multiply all the parameters $g_i$ and $\s$ that appear in \eqref{1.1} by a positive constant $a$, resulting in the new parameters
\begin{equation}\label{11.0}
ag_1,\ldots,ag_n,\quad\text{ and }\quad a^2\s^2,
\end{equation}
 then the Atlas model generated by the new parameters is $Y_i=aX_i$,  and we see  from \eqref{1} that 
\begin{equation}\label{11}
\V_{Y_{(1)}}(t)=a^2\V_{X_{(1)}}(t),\quad\text{ and }\quad \VV_{Y_{(1)}}(t)=\VV_{X_{(1)}}(t),
\end{equation}
for all $t$. Second, if we multiply the parameters $g_i$ by a positive constant $a$ and we multiply $\s$ by $\sqrt{a}$, resulting in the new parameters
\begin{equation}\label{12.0}
ag_1,\ldots,ag_n,\quad\text{ and }\quad a\s^2,
\end{equation}
then the Atlas model generated by the new parameters is a time-changed version of \eqref{1.1} that behaves like $X_i(at)$. In this case it follows from  \eqref{1} that
\begin{equation}\label{12}
\V_{Y_{(1)}}(t)=a\V_{X_{(1)}}(at),\quad\text{ and }\quad \VV_{Y_{(1)}}(t)=\VV_{X_{(1)}}(at),
\end{equation}
for all $t$.

A {\em simple} Atlas model is one with a single growth parameter $g>0$ such that parameters  $g_1=\cdots=g_{n-1}=-g$ and $g_n=(n-1)g$. In the case of a simple Atlas model,  by using successive transformations of the parameters of the form \eqref{11.0} and \eqref{12.0} it is possible to express the variogram in terms of the variogram of the canonical Atlas model of depth $n$ defined by
\begin{equation*}
dX_i(t) =  \big(n\1_{\{X_i(t)=X_{(n)}(t)\}}-1\big)dt + dW_i(t).
\end{equation*}
Hence, we have 

\vspace{5pt}
\noindent{\bf Proposition 1:}  {\em A simple Atlas model can be identified by the variogram of its top-ranked process.} 
\vspace{5pt}

To observe how this method might work in practice, we simulated a simple Atlas model with $n=10$ over 10~million iterations. The parameters for the model \eqref{1.1} were 
\begin{equation}\label{10}
g_1,\ldots,g_9=-.0001,\quad g_{10}=.0009, \quad\text{ and }\quad \s^2=0001.
\end{equation} 
The variogram was sampled at powers of 2, $t=1,2,4,\cdots,524288$, and the estimated relative variogram appears in Figure~\ref{f1} below. The value tends to about $1/10$, correctly indicating that the depth of the model is 10, but the 10 million iterations used to obtain this estimate suggest that this method is more of theoretical than practical interest. A second variogram was generated for the same model \eqref{1.1} with parameters
\begin{equation}\label{13}
g_1,\ldots,g_9=-.0002,\quad g_{10}=.0018, \quad\text{ and }\quad \s^2=0001,
\end{equation} 
and we see in Figure~\ref{f1} that the variogram for these new parameters, represented by the red dots, converges to 0.1 about four times more quickly than the variogram corresponding to the old parameters \eqref{10}, as is expected following \eqref{11} and \eqref{12}. 

This leads to some perhaps more interesting problems:

\vspace{10pt}
\noindent{\bf Problem 1:}  In the general case \eqref{1.1}, can all the parameters $g_1,\ldots,g_n$ be recovered from the values of $\V_{X_{(1)}}(t)$? 
\vspace{10pt}

\noindent{\bf Problem 2:}  Characterize  ---  in the manner of the law of the iterated logarithm --- the long-term behavior of $X_{(1)}$ for an infinite Atlas model.

\vspace*{-20pt}
\begin{figure}[H]
\begin{center}
\hspace*{-20pt}
\scalebox{.75}{ \rotatebox{0}{
\includegraphics{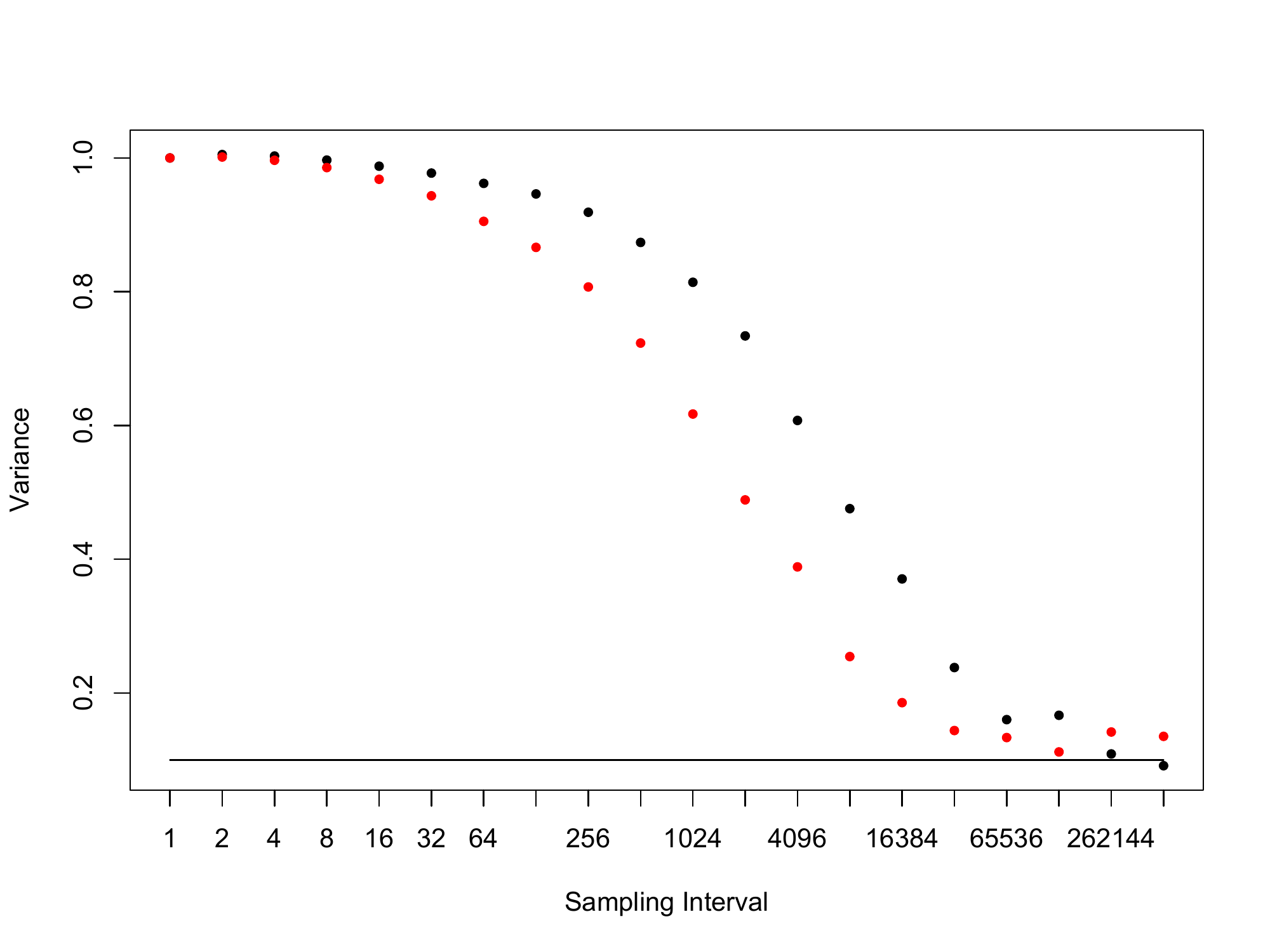}}}
\vspace{-10pt}
\caption{Estimated relative variograms $\VV_{X_{(1)}}$, with horizontal line at  0.1.}\label{f1}
\centerline{With parameters \eqref{10}, black dots; with parameters \eqref{13}, red dots.}
\end{center}
\end{figure}

\bibliographystyle{chicago}
\bibliography{math}

\end{document}